\documentclass[letterpaper, 10 pt, conference]{ieeeconf}

\IEEEoverridecommandlockouts                              
\usepackage{amsmath}    
\usepackage{amsfonts}
\usepackage{graphicx}   
\usepackage{subcaption}
\usepackage{epsfig} 
\usepackage{amssymb}
\usepackage{color}
\usepackage{cite}
\usepackage{float}

\newcommand{\R}{\mathbb{R}}
\newcommand{\cset}{\mathcal{U}}
\newcommand{\cfset}{\mathbb{U}}
\newcommand{\dfset}{\mathbb{D}}
\newcommand{\dset}{\mathcal{D}}
\newcommand{\obsset}{\mathcal{G}} 
\newcommand{\ioset}{\mathcal{O}} 
\newcommand{\brs}{\mathcal{V}} 
\newcommand{\frs}{\mathcal{W}} 
\newcommand{\pfrs}{\mathcal{P}} 
\newcommand{\targetset}{\mathcal{T}}

\newcommand{\ldt}{t^\text{LDT}}
\newcommand{\sta}{t^\text{STA}}
\newcommand{\dz}{\mathcal{A}} 
\newcommand{\cradius}{R_c} 
\newcommand{\pos}{p} 
\newcommand{\npos}{h} 
\newcommand{\veh}{Q} 
\newcommand{\dist}{\text{dist}} 
\newcommand{\rc}{R_c} 
\newcommand{\errorbound}{\mathcal{E}} 
\newcommand{\disckernel}{\Omega} 
\newcommand{\tracklaw}{\kappa} 
\newcommand{\state}{x} 
\newcommand{\ctrl}{u} 
\newcommand{\dstb}{d} 
\newcommand{\fdyn}{f} 
\newcommand{\costate}{\lambda}

\title{\large \bf
Safe Sequential Path Planning Under Disturbances and Imperfect Information}
\author{Somil Bansal*, Mo Chen*, Jaime F. Fisac, and Claire J. Tomlin
\thanks{This work has been supported in part by NSF under CPS:ActionWebs (CNS-931843), by ONR under the HUNT (N0014-08-0696) and SMARTS (N00014-09-1-1051) MURIs and by grant N00014-12-1-0609, by AFOSR under the CHASE MURI (FA9550-10-1-0567). The research of J. F. Fisac has received funding from the ``la Caixa" Foundation.}
\thanks{* Both authors contributed equally to this work. All authors are with the Department of Electrical Engineering and Computer Sciences, University of California, Berkeley. \{somil, mochen72, jfisac, tomlin\}@eecs.berkeley.edu}
}

\begin{document}
\maketitle
\thispagestyle{empty}
\pagestyle{empty}

\begin{abstract}
Multi-UAV systems are safety-critical, and guarantees must be made to ensure no unsafe configurations occur. Hamilton-Jacobi (HJ) reachability is ideal for analyzing such safety-critical systems; however, its direct application is limited to small-scale systems of no more than two vehicles due to an exponentially-scaling computational complexity. Previously, the sequential path planning (SPP) method, which assigns strict priorities to vehicles, was proposed; SPP allows multi-vehicle path planning to be done with a linearly-scaling computational complexity. However, the previous formulation assumed that there are no disturbances, and that every vehicle has perfect knowledge of higher-priority vehicles' positions. In this paper, we make SPP more practical by providing three different methods to account for disturbances in dynamics and imperfect knowledge of higher-priority vehicles' states. Each method has different assumptions about information sharing. We demonstrate our proposed methods in simulations.
\end{abstract}

\section{Introduction}
Recently, there has been an immense surge of interest in using unmanned aerial systems (UASs) for civil purposes \cite{Debusk10, Amazon16, AUVSI16, BBC16}. Many of these applications will involve unmanned aerial vehicles (UAVs) flying in urban environments. As a result, government agencies such as the Federal Aviation Administration (FAA) and National Aeronautics and Space Administration (NASA) of the United States are trying to develop new scalable ways to organize an air space in which potentially thousands of UAVs can fly together \cite{FAA13, Kopardekar16}.

One essential problem that needs to be addressed is how a group of vehicles in the same vicinity can reach their destinations while avoiding collision with each other. In some previous studies that address this problem, specific control strategies for the vehicles are assumed, and approaches such as induced velocity obstacles have been used \cite{Fiorini98, Chasparis05, Vandenberg08}. Other researchers have used ideas involving virtual potential fields to maintain collision avoidance while maintaining a specific formation \cite{Saber02, Chuang07}. Although interesting results emerge from these studies, simultaneous trajectory planning and collision avoidance were not considered. 

Trajectory planning and collision avoidance problems in safety-critical systems have been studied using Hamilton-Jacobi (HJ) reachability analysis, which provides guarantees on the success and safety of optimal system trajectories \cite{Barron90, Mitchell05, Bokanowski10, Margellos11, Fisac15}. In this context, one computes the reachable set, defined as the set of states from which the system can be driven to a target set. HJ reachability has been successfully used in applications involving systems with no more than two vehicles \cite{Mitchell05, Ding08, Huang11, Bayen07}. However, HJ reachability cannot be directly applied to systems involving multiple vehicles due to its exponentially scaling computational complexity. 

To overcome this problem, \cite{Chen15} presents sequential path planning (SPP), in which vehicles are assigned a strict priority ordering. In SPP, higher-priority vehicles ignore the lower-priority vehicles, which must take into account the presence of higher-priority vehicles by treating them as induced time-varying obstacles. Under this structure, computation complexity scales just \textit{linearly} with the number of vehicles. In addition, a structure like this has the potential to flexibly divide up the airspace for the use of many UAVs; this is an important task in NASA's concept of operations for UAS traffic management \cite{Kopardekar16}. 

The formulation in \cite{Chen15}, however, ignores disturbances and assumes perfect information about other vehicles' trajectories. In presence of disturbances, a vehicle's state trajectory evolution cannot be precisely known \textit{a priori}; thus, it is impossible to commit to exact trajectories as required in \cite{Chen15}. In such a scenario, a lower-priority vehicle needs to account for all possible states that the higher-priority vehicles could be in. To do this, the lower-priority vehicle needs to have some knowledge about the control policy used by each higher-priority vehicle. The main contribution of this paper is to take advantage of the computation benefits of the SPP scheme while resolving some of its practical challenges. In particular, we achieve the following:
\begin{itemize}
\item incorporate disturbances into the vehicle models,
\item analyze three different assumptions on information to which lower-priority vehicles may have access,
\item for each information pattern, propose a reachability-based method to compute the induced obstacles and the reachable sets that guarantee collision avoidance as well as successful transit to the destination.
\end{itemize}

\section{Problem Formulation \label{sec:formulation}}
Consider $N$ vehicles, denoted $\veh_i, i = 1,\ldots,n$, whose dynamics are described by the ordinary differential equation

\begin{equation}
\label{eq:dyn}
\begin{aligned}
\dot{x}_i &= f_i(t, x_i, u_i, d_i), \quad t \le \sta_i \\
u_i &\in \cset_i, d_i \in \dset_i, \quad i = 1,\ldots, N
\end{aligned}
\end{equation}

\noindent where $x_i \in \R^{n_i}, u_i$ denote the state and control of $i$th vehicle $\veh_i$ respectively, and $d_i$ denotes the disturbance experienced by $\veh_i$. In general, the physical meaning of $x_i$ and the dynamics $f_i$ depend on the specific dynamic model of $\veh_i$, and need not be the same across the different vehicles. $\sta_i$ in \eqref{eq:dyn} denotes the scheduled time of arrival of $\veh_i$. 

For convenience, we will use the sets $\cfset_i, \dfset_i$ to denote the set of functions from which the control and disturbance functions $u_i(\cdot), d_i(\cdot)$ can be drawn. 
Let $\pos_i \in \R^\pos$ denote the position of $\veh_i$. 
Denote the rest of the states $\npos_i$, so that $x_i = (\pos_i, \npos_i)$. The initial state of $\veh_i$ is given by $x_{i0}$. Under the worst case disturbance, each vehicle aims to get to some set of target states, denoted $\targetset_i \subset \R^{n_i}$, by some scheduled time of arrival $\sta_i$. On its way to $\targetset_i$, each vehicle must avoid the danger zones $\dz_{ij}(t)$ of all other vehicles $j\neq i$ for all time. In general, the danger zone can be defined to capture any undesirable configurations between $\veh_i$ and $\veh_j$. In this paper, we define $\dz_{ij}(t)$ as

\vspace{-0.5em} 
\begin{equation}
\dz_{ij}(t) = \{x_i \in \R^{n_i}: \|\pos_i - \pos_j(t)\|_2 \le \cradius \}
\end{equation}

\noindent the interpretation of which is that a vehicle is in another vehicle's danger zone if the two vehicles are within a Euclidean distance of $\cradius$ apart. 
 
The problem of driving each of the vehicles in \eqref{eq:dyn} into their respective target sets $\targetset_i$ would be in general a differential game of dimension $\sum_i n_i$. However, due to the exponential scaling of the complexity with the problem dimension, an optimal solution is computationally intractable even for $N>2$, with $n_i$ as small as $3$.

In this paper, we assume that vehicles have assigned priorities as in the SPP method \cite{Chen15}. Since the analysis in \cite{Chen15} did not take into account the presence of disturbances $d_i$ and limited information available to each vehicle, we extend the work in \cite{Chen15} to answer the following:
\begin{enumerate}
\item How can each vehicle guarantee that it will reach its target set without getting into any danger zones, despite the disturbances it and other vehicles experience?
\item How should each vehicle robustly handle situations with limited information about the state, control policy, and intention of other vehicles?
\end{enumerate}
%

\section{Background \label{sec:background}}
This section provides a brief summary of \cite{Chen15}, in which the SPP scheme is proposed under perfect information and absence of disturbances. Here, the dynamics of $\veh_i$ becomes

\begin{equation}
\label{eq:dyn_no_dstb}
\begin{aligned}
\dot{x}_i &= f_i(t, x_i, u_i), \quad t \le \sta_i \\
u_i &\in \cset_i, \quad i = 1,\ldots, N
\end{aligned}
\end{equation}

\noindent where the difference compared to \eqref{eq:dyn} is that the disturbance $d_i$ is no longer a part of the dynamics.

In order to make the $N$-vehicle path planning problem safe and tractable, a reasonable structure is imposed to the problem: the vehicles are assigned a strict priority ordering. When planning its trajectory to its target, a higher-priority vehicle can disregard the presence of a lower-priority vehicle. In contrast, a lower-priority vehicle must take into account the presence of all higher-priority vehicles, and plan its trajectory in a way that avoids the higher-priority vehicles' danger zones. For convenience and without lost of generality, let $\veh_i$ be the vehicle with the $i$th highest priority. 

Under the above convention, each vehicle $\veh_i$ must take into account time-varying obstacles induced by vehicles $\veh_j, j<i$, denoted $\ioset_i^j(t)$ and represent the set of states that could possibly be in the danger zone of $\veh_j$. Optimal safe path planning of each lower-priority vehicle $\veh_i$ then consists of determining the optimal path that allows $\veh_i$ to reach its target $\targetset_i$ while avoiding the time-varying obstacles $\obsset_i(t)$, defined by

\begin{equation}
\label{eq:ioset}
\obsset_i(t) = \bigcup_{j=1}^{i-1}\ioset_i^j(t)
\end{equation}

Such an optimal path planning problem can be solved by computing a backward reachable set (BRS) $\brs_i(t)$ from a target set $\targetset_i$ using formulations of HJ variational inequalities (VI) such as \cite{Barron90, Bokanowski10, Bokanowski11, Fisac15}. For example, to compute BRSs under the presence of time-varying obstacles, the authors in \cite{Bokanowski11} augment system with the time variable, and then applied reachability theory for time-invariant systems. To avoid increasing the problem dimension and save computation time, for the simulations of this paper we utilize the formulation in \cite{Fisac15}, which does not require augmentation of the state space with the time variable.

Starting from the highest-priority vehicle $\veh_1$, one computes the BRS $\brs_1(t)$, from which the optimal control and trajectory $x_1(\cdot)$ to $\targetset_1$ can be obtained. Under the absence of disturbances and perfect information, obstacles induced by a higher-priority vehicle $\veh_j$, starting with $j=1$, for a lower-priority vehicle $\veh_i$ is simply the danger zone centered around the position $p_j(\cdot)$ of each point on the trajectory:
\vspace{-0.5em} 
\begin{equation}
\ioset_i^j(t) = \{x_i: \|p_i - p_j(t)\|\le\cradius\}
\end{equation}

Given $\ioset_i^j(t), j<i$, and continuing with $i = 2$, the optimal safe trajectories for each vehicle $\veh_i$ can be computed. All of the trajectories are optimal in the sense that given the requirement that $\veh_i$ must arrive at $\targetset_i$ by time $\sta_i$, the latest departure time $\ldt_i$ and the optimal control $u^*_i(\cdot)$ that guarantees arrival by $\sta_i$ can be obtained.

To compute $\brs_i(t)$ using the method in \cite{Fisac15}, we solve the following HJ VI for $t\le\sta_i$:
\vspace{-0.5em} 
\begin{equation}
\label{eq:HJIVI}
\begin{aligned}
\max\Big\{&\min\big\{D_t V_i(t, x_i) + H_i\left(t, x_i, D_{x_i} V_i\right), \\
&l_i(x_i) - V_i(t, x_i)\big\}, -g_i(t, x_i) - V_i(t, x_i)\Big\} = 0 \\
&V_i(\sta_i, x_i) = \max\big\{l_i(x_i), -g_i(0, x_i)\big\}\\ 
\end{aligned}
\end{equation}
\begin{equation}
\label{eq:basicSPPHam}
H_i\left(t, x_i, \lambda\right) = \min_{u_i \in \cset_i} \lambda \cdot f_i(t, x_i, u_i)
\end{equation}

\noindent where $\lambda$ is the gradient of the value function, $D_{x_i} V_i$, and $l_i(x_i), g_i(t,x_i),V_i(t,x_i)$ are implicit surface functions representing the target $\targetset_i$, the time-varying obstacles $\obsset_i(t)$, and the backward reachable set $\brs_i(t)$, respectively: 
\begin{equation}
\label{eq:impl_surf}
\begin{aligned}
x_i \in \targetset_i &\Leftrightarrow l_i(x_i) \le 0 \\
x_i(t) \in \obsset_i(t) &\Leftrightarrow g_i(t,x_i) \le 0 \\
x_i(t) \in \brs_i(t) &\Leftrightarrow V_i(t, x_i) \le 0
\end{aligned}
\end{equation}

The optimal control is given by
 \vspace{-0.4em} 
\begin{equation}
\label{eq:optCtrl}
u^*_i\left(t, x_i\right) = \arg \min_{u_i \in \cset_i} \lambda \cdot f_i(t, x_i, u_i)
\end{equation}

\section{Disturbances and Incomplete Information \label{sec:obs_gen}}
Disturbances and incomplete information significantly complicate the SPP scheme. The main difference is that the vehicle dynamics satisfy \eqref{eq:dyn} as opposed to \eqref{eq:dyn_no_dstb}. Committing to exact trajectories is therefore no longer possible, since the disturbance $d_i(\cdot)$ is \textit{a priori} unknown. Thus, the induced obstacles $\ioset_i^j(t)$ are no longer just the danger zones centered around positions. We present three methods to address the above issues. The methods differ in terms of control policy information that is known to a lower-priority vehicle about a higher-priority vehicle, and have their relative advantages and disadvantages depending on the situation. The three methods are as follows:
\begin{itemize}
\item \textbf{Centralized control}: A specific control strategy is enforced upon a vehicle; this can be achieved, for example, by some central agent such as an air traffic controller.
\item \textbf{Least restrictive control}: A vehicle is required to arrive at its targets on time, but has no other restrictions.
\item \textbf{Robust trajectory tracking}: A vehicle declares a nominal trajectory which can be robustly tracked.
\end{itemize}

In general, the above methods can be used in combination in a single path planning problem, with each vehicle independently having different control policies. Lower-priority vehicles would then plan their paths while taking into account the control policy information known for each higher-priority vehicle. For clarity, we will present each method as if all vehicles are using the same method of path planning.

For simplicity of explanation, we assume that no static obstacles exist. If static obstacles do exist, the time-varying obstacles $\obsset_i(t)$ simply become the union of the induced obstacles $\ioset_i^j(t)$ in \eqref{eq:ioset} and the static obstacles.

\subsection{Method 1: Centralized Control \label{sec:cc}}
The highest-priority vehicle $\veh_1$ first plans its path by computing the BRS (with $i=1$)
\vspace{-0.3em}
\begin{equation}
\label{eq:BRS}
\begin{aligned}
&\brs_i(t) = \{x_i: \exists u_i(\cdot) \in \cfset_i, \forall d_i(\cdot) \in \dfset_i, x_i(\cdot) \text{ satisfies \eqref{eq:dyn}}, \\
&\forall s \in [t, \sta_i], x_i(s) \notin \obsset_i(s), \exists s \in [t, \sta_i], x_i(s) \in \targetset_i\}
\end{aligned}
\end{equation}

Since we have assumed no static obstacles exist, we have that for $\veh_1, \obsset_1(s)=\emptyset ~ \forall s \le \sta_i$, and thus the above BRS is well-defined. This BRS can be computed by solving the HJ VI \eqref{eq:HJIVI} with the following Hamiltonian:
\vspace{-0.3em}
\begin{equation}
H_i\left(t, x_i, \lambda\right) = \min_{u_i \in \cset_i} \max_{d_i \in \dset_i} \lambda \cdot f_i(t, x_i, u_i, d_i)
\end{equation}

\noindent where $l_i(x_i), g_i(t,x_i),V_i(t,x_i)$ are implicit surface functions representing the target $\targetset_i, \obsset_i(t), \brs_i(t)$, respectively. From the BRS, we can obtain the optimal control
\vspace{-0.3em}
\begin{equation}
\label{eq:opt_ctrl_i}
u_i^*(t,x_i) =  \arg \min_{u_i \in \cset_i} \max_{d_i \in \dset_i} \lambda \cdot f_i(t, x_i, u_i, d_i)
\end{equation}

Here, as well as in the other two methods, the latest departure time $\ldt_i$ is then given by $\arg \sup_t x_{i0} \in \brs_i(t)$.

If there is a centralized controller directly controlling each of the $N$ vehicles, then the control law of each vehicle can be enforced. In this case, lower-priority vehicles can safely assume that higher-priority vehicles are applying the enforced control law. In particular, the optimal controller for getting to the target, $u^*_i(t, x_i)$ can be enforced. In this case, the dynamics of each vehicle becomes 
\vspace{-0.3em}
\begin{equation}
\label{eq:dyn_cc}
\begin{aligned}
\dot x_i &= f^*_i (t, x_i, d_i) = f_i(t, x_i, u^*_i(t,x_i), d_i) \\
d_i &\in \dset_i, \quad i = 1,\ldots, N, \quad t \in [\ldt_i, \sta_i]
\end{aligned}
\end{equation}

\noindent where $u_i$ no longer appears explicitly in the dynamics.

From the perspective of a lower-priority vehicle $\veh_i$, a higher-priority vehicle $\veh_j, j < i$ induces a time-varying obstacle that represents the positions that could possibly be within the capture radius $\cradius$ of $\veh_j$ under the dynamics $f^*_j(t, x_j, d_j)$. Determining this obstacle involves computing a forward reachable set (FRS) of $\veh_j$ starting from $x_j(\ldt_j) = x_{j0}$. The FRS $\frs_j(t)$ is defined as follows:
\vspace{-0.3em}
\begin{equation}
\label{eq:FRS1}
\begin{aligned}
&\frs_j(t) = \{y \in \R^{n_j}: \exists d_j(\cdot) \in \dfset_j, \\
&x_j(\cdot) \text{ satisfies \eqref{eq:dyn_cc}}, x_j(\ldt_j) = x_{j0}, x_j(t) = y\}
\end{aligned}
\end{equation}

The FRS can be computed using the following HJ VI:
\vspace{-0.4em}
\begin{equation}
\label{eq:FRS_HJIVI}
\begin{aligned}
&D_t W_j(t, x_j) + H_j\left(t, x_j, D_{x_j} W_j\right) = 0, t \in [\ldt_j, \sta_j]\\
&\quad W_j(\ldt_j, x_j) = \bar l_j(x_j) \\
&\quad H_j\left(t, x_j, \lambda\right) = \max_{d_j \in \dset_j} \lambda \cdot f^*_j(t, x_j, d_j)
\end{aligned}
\end{equation}

\noindent where $\bar l$ is chosen to be\footnote{In practice, we define the target set to be a small region around the vehicle's initial state for computational reasons.} such that $\bar l (y) = 0 \Leftrightarrow y = x_j(\ldt_j)$.

The FRS $\frs_j(t)$ represents the set of possible states at time $t$ of a higher-priority vehicle $\veh_j$ given all possible disturbances $d_j(\cdot)$ and given that $\veh_j$ uses the feedback controller $u_j^*(t, x_j)$. In order for a lower-priority vehicle $\veh_i$ to guarantee that it does not go within a distance of $\cradius$ to $\veh_j$, $\veh_i$ must stay a distance of at least $\cradius$ away from the set $\frs_j(t)$ for all possible values of the non-position states $\npos_j$. This gives the obstacle induced by a higher-priority vehicle $\veh_j$ for a lower-priority vehicle $\veh_i$ as follows:
\vspace{-0.4em}
\begin{equation}
\ioset_i^j(t) = \{x_i: \dist(\pos_i, \pfrs_j(t)) \le \cradius \}
\end{equation}

\noindent where the $\dist(\cdot, \cdot)$ function represents the minimum distance from a point to a set, and the set $\pfrs_j(t)$ is the set of states in the FRS $\frs_j(t)$ projected onto the states representing position $\pos_j$, and disregarding the non-position dimensions $\npos_j$:
\vspace{-0.4em}
\begin{equation}
\pfrs_j(t) = \{p_j: \exists \npos_j, (p_j, \npos_j) \in \frs_j(t)\}.
\end{equation}

Finally, taking the union of the induced obstacles $\ioset_i^j(t)$ as in \eqref{eq:ioset} gives us the time-varying obstacles $\obsset_i(t)$ needed to define and determine the BRS $\brs_i(t)$ in \eqref{eq:BRS}. Repeating this process, all vehicles will be able to plan paths that guarantee the vehicles' timely and safe arrival.
\subsection{Method 2: Least Restrictive Control \label{sec:lrc}}
Here, we again begin with the highest-priority vehicle $\veh_1$ planning its path by computing the BRS $\brs_i(t)$ in \eqref{eq:BRS}. However, if there is no centralized controller to enforce the control policy for higher-priority vehicles, weaker assumptions must be made by the lower-priority vehicles to ensure collision avoidance. One reasonable assumption that a lower-priority vehicle can make is that all higher-priority vehicles follow the least restrictive control that would take them to their targets. This control would be given by 
\vspace{-0.4em}
\begin{equation}
\label{eq:lrctrl} 
u_j(t, x_j)\in \begin{cases} \{u_j^*(t, x_j) \text{ given by } \eqref{eq:opt_ctrl_i}\} \text{ if } x_j(t)\in \partial \brs_j(t), \\
\cset_j  \text{ otherwise}
\end{cases}
\end{equation}

Such a controller allows each vehicle to use any controller, except when it is on the boundary of the BRS, $\partial \brs_j(t)$, in which case $u_j^*(t, x_j)$ given by \eqref{eq:opt_ctrl_i} must be used to get to the target safely and on time. This assumption is the weakest one that could be made by lower-priority vehicles given that the higher-priority vehicles will get to their targets on time.

Suppose a lower-priority vehicle $\veh_i$ assumes that higher-priority vehicles $\veh_j, j < i$ use the least restrictive control strategy in \eqref{eq:lrctrl}. From the perspective of $\veh_i$, a higher-priority vehicle $\veh_j$ could be in any state that is reachable from $\veh_j$'s initial state $x_j(\ldt_j) = x_{j0}$ and from which the target $\targetset_j$ can be reached. Mathematically, this is defined by the intersection of a FRS from the initial state $x_j(\ldt)=x_{j0}$ and the BRS defined in \eqref{eq:BRS} from the target set $\targetset_j$, $\brs_j(t) \cap \frs_j(t)$. In this situation, since $\veh_j$ cannot be assumed to be using any particular feedback control, $\frs_j(t)$ is defined as
\vspace{-0.4em}

\begin{equation}
\label{eq:FRS2}
\begin{aligned}
&\frs_j(t) = \{y \in \R^{n_j}: \exists u_j(\cdot)\in\cfset_j, \exists d_j(\cdot) \in \dfset_j, \\
&x_j(\cdot) \text{ satisfies \eqref{eq:dyn}}, x_j(\ldt_j) = x_{j0}, x_j(t) = y\}
\end{aligned}
\end{equation}

This FRS can be computed by solving \eqref{eq:FRS_HJIVI} without obstacles, and with
\vspace{-0.4em}
\begin{equation}
H_j\left(t, x_j, \lambda\right) = \max_{u_j \in \cset_j} \max_{d_j \in \dset_j} \lambda \cdot f_j(t, x_j, u_j, d_j)
\end{equation}

In turn, the obstacle induced by a higher-priority $\veh_j$ for a lower-priority vehicle $\veh_i$ is as follows:
\vspace{-0.4em}
\begin{equation}
\begin{aligned}
\ioset_i^j(t) &= \{x_i: \dist(\pos_i, \pfrs_j(t)) \le \cradius \}, \text{ with} \\
\pfrs_j(t) &= \{p_j: \exists \npos_j, (p_j, \npos_j) \in \brs_j(t) \cap \frs_j(t)\}
\end{aligned}
\end{equation}
%
\subsection{Method 3: Robust Trajectory Tracking\label{sec:rtt}}
Although it is impossible to commit to and track an exact trajectory in the presence of disturbances, it may still be possible to \textit{robustly} track a \textit{nominal} trajectory with a bounded error at all times. If this can be done, then the tracking error bound can be used to determine the induced obstacles. Here, computation is done in two phases: the \textit{planning phase} and the \textit{disturbance rejection phase}. In the planning phase, we compute a nominal trajectory $\state_{r,j}(\cdot)$ that is feasible in the absence of disturbances. In the disturbance rejection phase, we compute a bound on the tracking error.

In the planning phase, planning is done for a reduced control set $\cset^p\subset\cset$, as some margin is needed to reject unexpected disturbances while tracking the nominal trajectory. In the disturbance rejection phase, we determine the error bound independently of the nominal trajectory. Let $\state_j$ and $\state_{r,j}$ denote the states of the actual vehicle $\veh_j$ and an arbitrary nominal trajectory, respectively, and define the tracking error $e_j=\state_j-\state_{r,j}$. When the error dynamics are independent of the absolute state as in \eqref{eq:edyn} (and also (7) in \cite{Mitchell05}), we can obtain error dynamics of the form
\begin{equation}
\label{eq:edyn} 
\begin{aligned}
\dot{e_j} &= \fdyn_{e_j}(e_j, \ctrl_j, \ctrl_{r,j},\dstb_j), \\
\ctrl_j &\in \cset_j, \ctrl_{r,j} \in \cset^p_j, \dstb_j \in \dset_j, \quad t \leq 0
\end{aligned}
\end{equation}

To obtain bounds on the tracking error, we first conservatively estimate the error bound around any reference state $\state_{r,j}$, denoted $\errorbound_j = \{e_j: \|\pos_{e_j}\|_2 \le R_{\text{EB}}\}$,
\noindent where $\pos_{e_j}$ denotes the position coordinates of $e_j$ and $R_{\text{EB}}$ is a design parameter. We next solve a reachability problem with its complement $\errorbound_j^c$, the set of tracking errors violating the error bound, as the target in the space of the error dynamics. From $\errorbound_j^c$, we compute the following BRS:
\begin{equation} \label{eqn:errBound}
\begin{aligned}
&\brs^{\text{EB}}_{j}(t, 0) = \{y: \forall \ctrl_j(\cdot) \in \cfset_j, \exists \ctrl_{r, j}(\cdot) \in \cfset^\pos_j, \exists \dstb_j(\cdot) \in \dfset_i, \\
& e_j(\cdot) \text{ satisfies \eqref{eq:edyn}}, e_j(t) = y, \exists s \in [t, 0], e_j(s) \in \errorbound_j^c\}, 
\end{aligned}
\end{equation}
where the Hamiltonian to compute the BRS is given by:
\begin{equation}
\begin{aligned}
H^{\text{EB}}_{j}(e_j, \costate) &= \max_{\ctrl_j \in \cset_j} \min_{\ctrl_{r, j} \in \cset^\pos_j, \dstb_j \in \dset_j} \costate \cdot \fdyn_{e_j}(e_j, \ctrl_j, \ctrl_{r,j}, \dstb_j).
\end{aligned}
\end{equation}

Letting $t \to -\infty$, we obtain the infinite-horizon control-invariant set $\disckernel_j := \lim_{t \to -\infty} \left( \brs^{\text{EB}}_{j}(t, 0) \right)^c$. If $\disckernel_j$ is nonempty, then the tracking error $e_j$ at flight time is guaranteed to remain within $\disckernel_j \subseteq \errorbound_j$ provided that the vehicle starts inside $\disckernel_j$ and subsequently applies the feedback control law
\begin{equation}
\label{eq:robust_tracking_law}
\tracklaw_j(e_j) = \arg\max_{\ctrl_j \in \cset_j} \min_{\ctrl_{r, j} \in\cset^\pos_j, \dstb_j \in \dset_j} \costate \cdot \fdyn_{e_j}(e_j,\ctrl_j,\ctrl_{r, j},\dstb_j).
\end{equation}

The induced obstacles by each higher-priority vehicle $\veh_j$ can thus be obtained by: 
\begin{equation} 
\label{eqn:rttObs}
\begin{aligned}
\ioset_i^j(t) &=  \{\state_i: \exists y \in \pfrs_j(t), \|\pos_i - y\|_2 \le \rc \} \\
\pfrs_j(t) &= \{\pos_j: \exists \npos_j, (\pos_j, \npos_j) \in \disckernel_j  + \state_{r,j}(t)\},
\end{aligned}
\end{equation}
\noindent where the ``$+$'' in \eqref{eqn:rttObs} denotes the Minkowski sum. 

Since each vehicle $\veh_j$, $j<i$, can only be guaranteed to stay within $\disckernel_j$, we must make sure during the path planning of $\veh_i$ that at any given time, the error bounds of $\veh_i$ and $\veh_j$, $\disckernel_i$ and $\disckernel_j$, do not intersect. This can be done by augmenting the total obstacle set by $\disckernel_i$:

\vspace{-1em}
\begin{equation} 
\label{eqn:rttAugObs}
\tilde{\obsset}_i(t) = \obsset_i(t) + \disckernel_i.
\vspace{-1em}
\end{equation}

Finally, given $\disckernel_i$, we can guarantee that $\veh_i$ will reach its target $\targetset_i$ if $\disckernel_i \subseteq \targetset_i$; thus, in the path planning phase, we modify $\targetset_i$ to be $\tilde{\targetset}_i := \{\state_i: \disckernel_i + \state_i \subseteq \targetset_i\}$, and compute a BRS, with the control authority $\cset^\pos_i$, that contains the initial state of the vehicle. Mathematically,

\vspace{-1.25em}
\begin{equation}
\label{eq:rttBRS}
\begin{aligned}
\brs_i^\text{rtt}(t, \sta_i) = & \{y: \exists \ctrl_i(\cdot) \in \cfset^p_i, \state_i(\cdot) \text{ satisfies \eqref{eq:dyn_no_dstb}},\\
&\forall s \in [t, \sta_i], \state_i(s) \notin \tilde{\obsset}_i(s), \\
& \exists s \in [t, \sta_i], \state_i(s) \in \tilde{\targetset}_i, \state_i(t) = y\}
\end{aligned}
\end{equation}

The Hamiltonian to compute $\brs_i^\text{rtt}(t, \sta_i)$ and the optimal control for reaching $\tilde{\targetset}_i$ are given by \eqref{eq:basicSPPHam} and \eqref{eq:optCtrl} respectively.
%
The nominal trajectory $\state_{r,i}(\cdot)$ can thus be obtained by using vehicle dynamics \eqref{eq:dyn_no_dstb}, with the optimal control  $\ctrl_i^\text{rtt}(\cdot)$. From the resulting nominal trajectory $\state_{r,i}(\cdot)$, the overall control policy to reach $\targetset_i$ can be obtained via \eqref{eq:robust_tracking_law}. 

\section{Numerical Simulations \label{sec:sim}}
We demonstrate our proposed methods using a four-vehicle example. Each vehicle has the following model:
\vspace{-0.2em}
\begin{equation*}
\label{eq:dyn_i}
\begin{aligned}
\dot{\pos}_{x,i} &= v_i \cos \theta_i + d_{x,i} &\\
\dot{\pos}_{y,i} &= v_i \sin \theta_i + d_{y,i} & \underline{v} \le v_i \le \bar{v}, |\omega_i| \le \bar{\omega}\\
\dot{\theta}_i &= \omega_i + d_{\theta,i} & \|(d_{x,i}, d_{y,i}) \|_2 \le d_{r}, |d_{\theta,i}| \le \bar{d_{\theta}}\\
\end{aligned}
\end{equation*}

\noindent where $p_i = (p_{x,i}, p_{y,i}), \theta_i, d = (d_{x,i}, d_{y,i}, d_{\theta,i})$ respectively represent $\veh_i$'s position, heading, and disturbances in the three states. The control of $\veh_i$ is $u_i = (v_i, \omega_i)$, where $v_i$ is the speed of $\veh_i$ and $\omega_i$ is the turn rate; both controls have a lower and upper bound. For illustration purposes, we choose $\underline{v} = 0.5, \bar{v} = 1, \bar\omega = 1$; however, our method can easily handle the case in which these inputs differ across vehicles and cases in which each vehicle has a different dynamic model. The disturbance bounds are chosen as $d_{r} = 0.1, \bar{d_{\theta}} = 0.2$, which correspond to a 10\% uncertainty in the dynamics. 

%

The initial states of the vehicles are given as follows:
\begin{equation}
\begin{aligned}
x_1^0 &= (-0.5, 0, 0), \quad &x_2^0 = (0.5, 0, \pi), \\
x_3^0 &= \left(-0.6, 0.6, 7\pi/4\right), \quad &x_4^0 = \left(0.6, 0.6, 5\pi/4\right).
\end{aligned}
\end{equation}

\noindent Each of the vehicles has a target set $\targetset_i$ that is circular in their position $\pos_i$ centered at $c_i = (c_{x,i}, c_{y,i})$ with radius $r$:
\vspace{-0.2em}
\begin{equation}
\targetset_i = \{x_i \in \R^3: \|p_i - c_i\| \le r\}
\end{equation}

\begin{figure}[t!]
\centering
\begin{subfigure}{.35\columnwidth}
  \centering
  \includegraphics[width=\columnwidth]{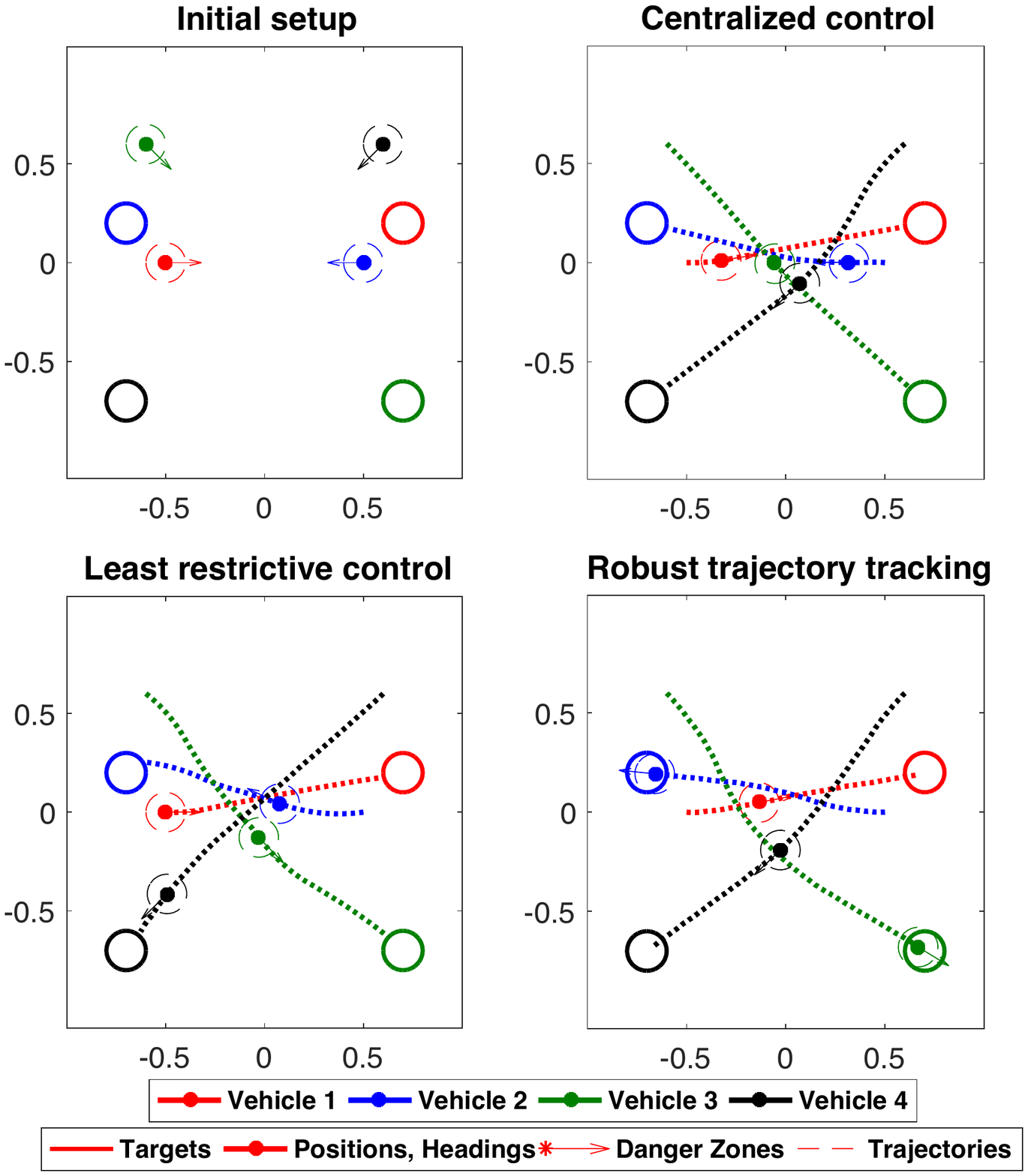}
  \subcaption{}
  \label{fig:initSetup}
\end{subfigure}%
\begin{subfigure}{.35\columnwidth}
  \centering
  \includegraphics[width=\columnwidth]{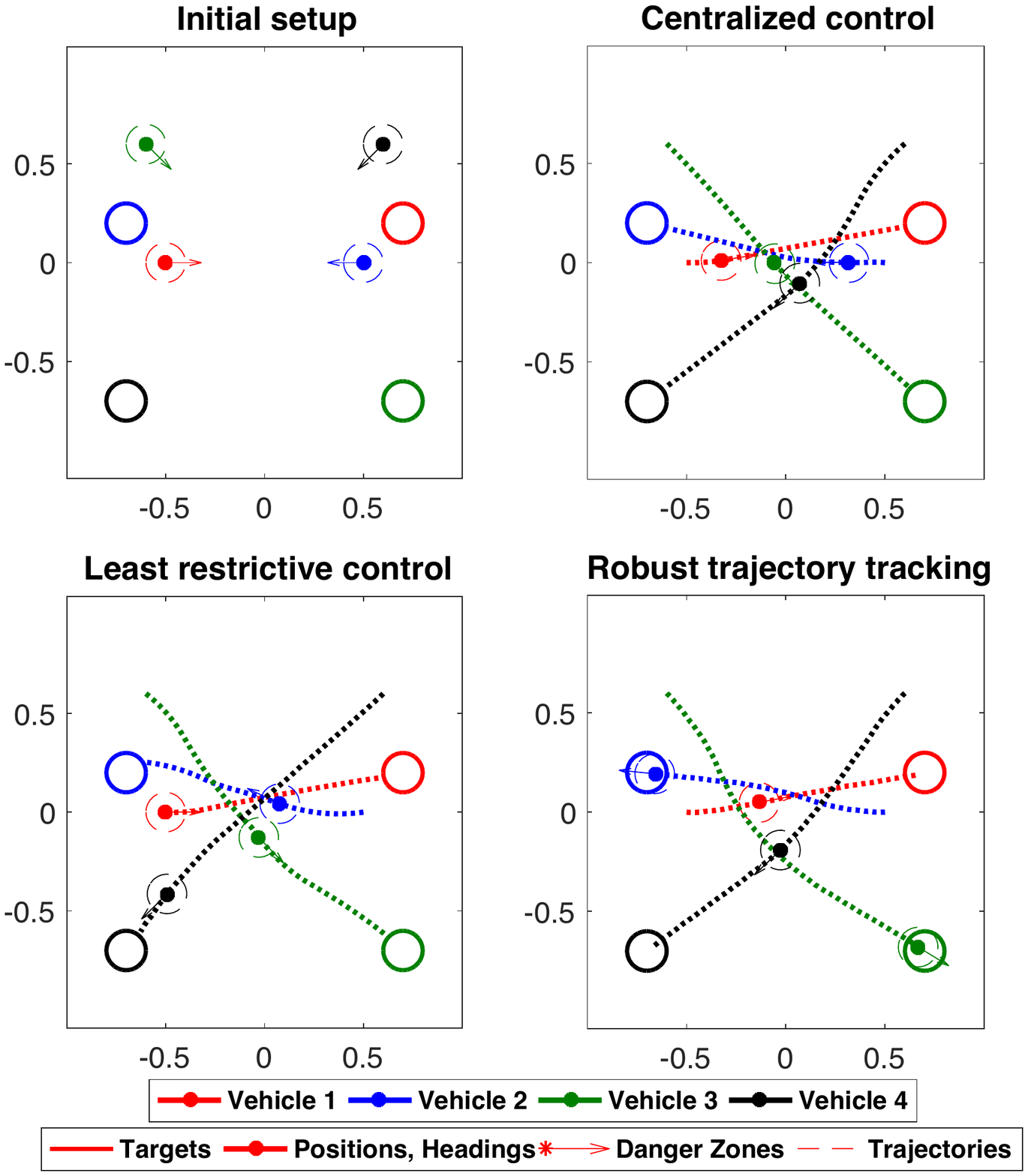}
  \subcaption{}
  \label{fig:trajsCC}
\end{subfigure}
\begin{subfigure}{.35\columnwidth}
  \centering
  \includegraphics[width=\columnwidth]{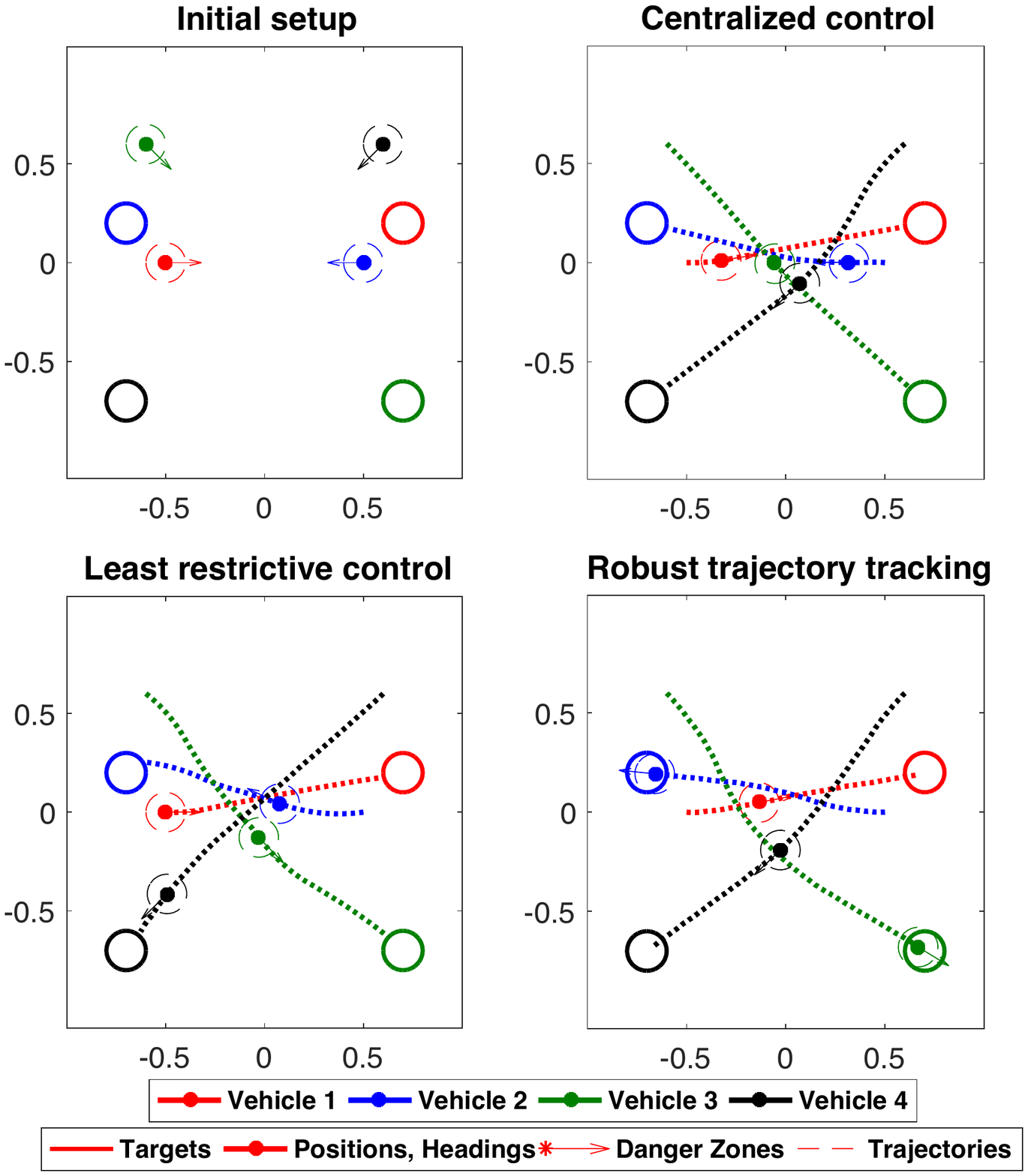}
  \subcaption{}
  \label{fig:trajsLRC}
\end{subfigure}
\begin{subfigure}{.35\columnwidth}
  \centering
  \includegraphics[width=\columnwidth]{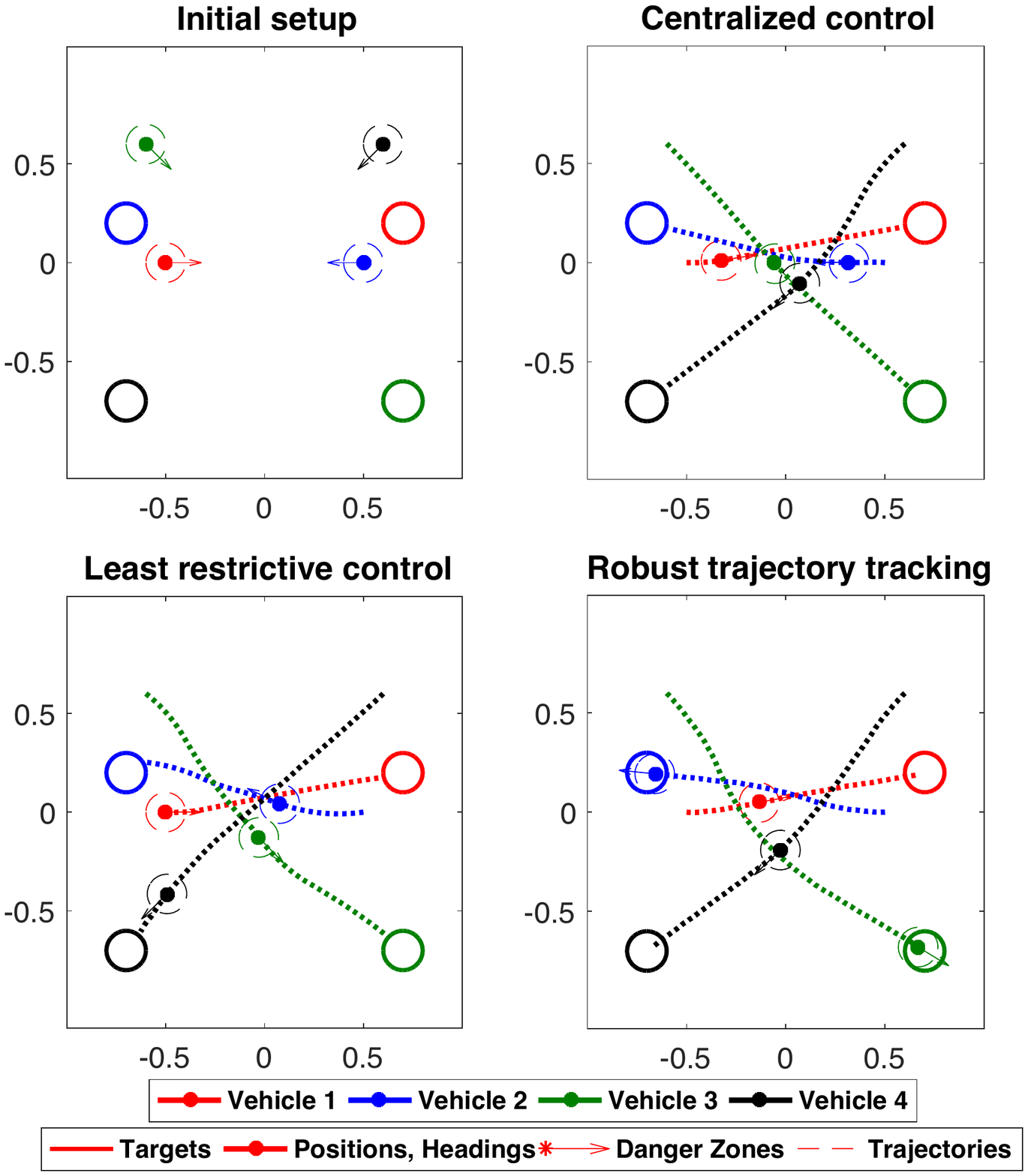}
  \subcaption{}
  \label{fig:trajsRTT}
\end{subfigure}
\begin{subfigure}{.7\columnwidth}
  \centering
  \includegraphics[width=\columnwidth]{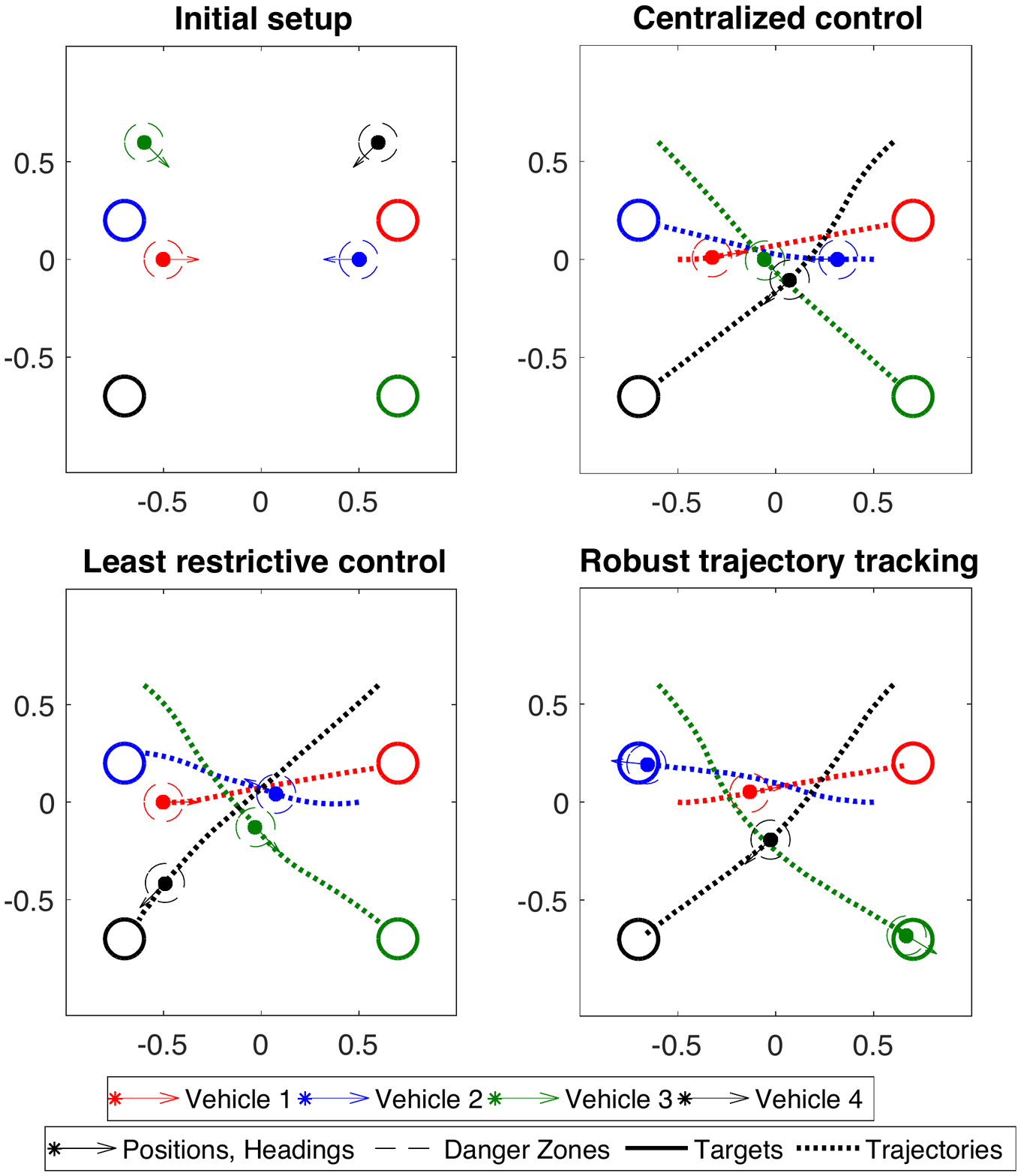}
\end{subfigure}%
  \caption{Initial configuration and simulated trajectories of the vehicles for the three proposed methods.}
  \label{fig:allTrajs}
\end{figure}

\noindent For the example shown, we chose $c_1 = (0.7, 0.2), c_2 = (-0.7, 0.2), c_3 = (0.7, -0.7), c_4 = (-0.7, -0.7)$ and $r = 0.1$. The setup of the example is shown in Fig. \ref{fig:initSetup}.

Using the SPP algorithms presented, we obtain $\ldt_i, i=1,2,3,4$ assuming $\sta_i=0$. Note that even though $\sta_i$ is assumed to be same for all vehicles in this example for simplicity, our method can easily handle the case in which $\sta_i$ is different for each vehicle.

For each proposed method of computing induced obstacles, we show the vehicles' entire trajectories (colored dotted lines), and overlay their positions (colored asterisks) and headings (arrows) at a point in time in which they are in relatively dense configuration. In all cases, the vehicles are able to avoid each other's danger zones (colored dashed circles) while getting to their target sets in minimum time. In addition, we show the evolution of the BRS over time for $\veh_3$ (green boundaries) as well as the obstacles induced by the higher-priority vehicles (black boundaries).

\begin{figure}[H]
  \centering
  \includegraphics[width=0.4\textwidth]{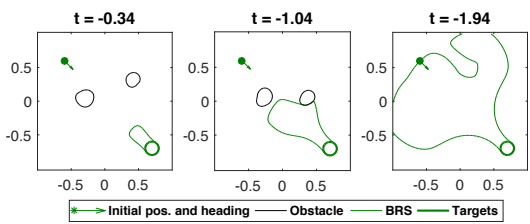}
  \caption{Evolution of the BRS and the obstacles induced by $\veh_1$ and $\veh_2$ for $\veh_3$ in the centralized control method.}
  \label{fig:cc_rs3}
\end{figure}
%
Fig. \ref{fig:trajsCC} shows simulated trajectories in the situation where each vehicle uses $u^*_i(t, x_i)$ in \eqref{eq:opt_ctrl_i}. In this case, vehicles appear to deviate slightly from a straight line trajectory towards their targets, just enough to avoid higher-priority vehicles. The deviation is small since the centralized controller is quite restrictive, making the possible positions of higher-priority vehicles cover a small area. In the dense configuration at $t=-1.0$, the vehicles are close to each other but still outside each other's danger zones.

Fig. \ref{fig:cc_rs3} shows the evolution of the BRS for $\veh_3$ (green boundary), as well as the obstacles (black boundary) induced by the higher-priority vehicles. The size of the obstacles remains relatively small. $\ldt_i$ numbers for the four vehicles (in order) in this case are $-1.35, -1.37, -1.94$ and $-2.04$. They are relatively close for the vehicles, because the obstacles generated by higher-priority vehicles are small and hence do not affect $\ldt$ of the lower-priority vehicles significantly. 
\subsection{Least Restrictive Control}
\begin{figure}[H]
  \centering
  \includegraphics[width=0.4\textwidth]{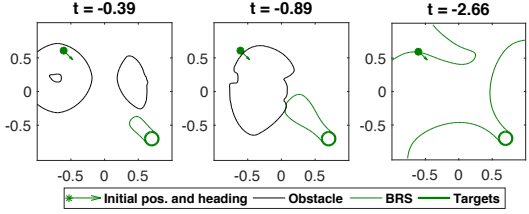}
  \caption{Evolution of the BRS for $\veh_3$ in the least restrictive control method. $\ldt_3$ is significantly lower than that in the centralized control method ($-1.94$ vs. $-2.66$).}
  \label{fig:lrc_rs3}
    \vspace{-1em}
\end{figure}

Fig. \ref{fig:trajsLRC} shows the simulated trajectories in the situation where each vehicle assumes that higher-priority vehicles use the least restrictive control to reach their targets, as described in \ref{sec:lrc}. Fig. \ref{fig:lrc_rs3} shows the BRS and induced obstacles for $\veh_3$.

$\veh_1$ (red) takes a relatively straight path to reach its target. From the perspective of all other vehicles, large obstacles are induced, since lower-priority vehicles make the weak assumption that higher-priority vehicles are using the least restrictive control. Because the obstacles induced are so large, it is optimal for lower-priority vehicles to wait until higher-priority vehicles pass. As a result, a dense configuration is never formed, and trajectories are relatively straight. The $\ldt_i$ values for vehicles are $-1.35, -1.97, -2.66$ and $-3.39$. Compared to the centralized control method, $\ldt_i$'s decrease significantly except for $\veh_1$, which need not account for any moving obstacles. 

From $\veh_3$'s (green) perspective, the large obstacles induced by $\veh_1$ and $\veh_2$ are shown in Fig. \ref{fig:lrc_rs3} as the black boundaries. As the BRS (green boundary) evolves over time, its growth gets inhibited by the large obstacles for a long time, as evident at $t=-0.89$. Eventually, the boundary of the BRS reaches the initial state of $\veh_3$ at $t = \ldt_3 = -2.66$.
%
%
\subsection{Robust Trajectory Tracking}
In the planning phase, we reduced the maximum turn rate of the vehicles from $1$ to $0.6$, and the speed range from $[0.5, 1]$ to exactly $0.75$ (constant speed). With these reduced control authorities, we determined from the disturbance rejection phase that any nominal trajectory from the planning phase can be robustly tracked within a distance of $0.075$.

\begin{figure}[H]
  \centering
  \includegraphics[width=0.4\textwidth]{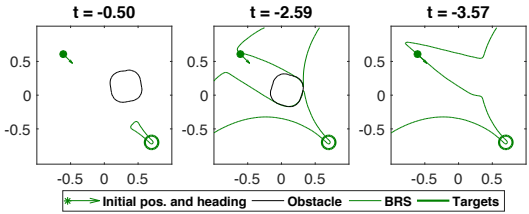}
  \caption{Evolution of the BRS for $\veh_3$ in the robust trajectory tracking method. Note that a smaller target set is used to ensure target reaching for any allowed tracking error.}
  \label{fig:rtt_rs3}
    \vspace{-1em}
\end{figure}

Fig. \ref{fig:trajsRTT} shows vehicle trajectories in the situation where each vehicle robustly tracks a nominal trajectory. Fig. \ref{fig:rtt_rs3} shows the BRS evolution and induced obstacles for $\veh_3$. 
%
%

In this case, the $\ldt_i$ values for the four vehicles are $-1.61, -3.16, -3.57$ and $-2.47$ respectively. In this method, vehicles use reduced control authority for path planning towards a reduced-size effective target set. As a result, higher-priority vehicles tend to have lower $\ldt$ compared to the other two methods, as evident from $\ldt_1$. Because of this ``sacrifice" made by the higher-priority vehicles during the path planning phase, the $\ldt$'s of lower-priority vehicles may increase compared to those in the other methods, as evident from $\ldt_4$. Overall, it is unclear how $\ldt_i$ will change for a vehicle compared to the other methods, as the conservative path planning increases $\ldt_i$ for higher-priority vehicles and decreases $\ldt_i$ for lower-priority vehicles.

\section{Conclusions}
We have proposed three different methods to account for disturbances and imperfect control policy information in sequential path planning; these three methods can be used independently across the different vehicles in the path planning problem. In each method, different assumptions about the control strategy of higher-priority vehicles are made. In all of the methods, all vehicles are guaranteed to successfully reach their respective destinations without entering each other's danger zones despite the worst-case disturbance the vehicles could experience. 

                                  
\bibliographystyle{IEEEtran}
\bibliography{references}
\end{document}